\documentclass[11pt]{article}

\renewcommand{\baselinestretch}{1.2}
  \renewcommand{\arraystretch}{1.0}
\usepackage{cite}
 \usepackage{amsmath}
\usepackage{amssymb}
\usepackage{graphicx}
\usepackage{ulem}
  \usepackage{hyperref}

\begin{document}
 \title{The Barth-Boneh-Waters Private Broadcast\\ Encryption Scheme Revisited}

 \author{Zhengjun Cao$^{1}$, \qquad Lihua Liu$^{2,*}$}
  \footnotetext{ $^1$Department of Mathematics, Shanghai University, Shanghai,
  China. \quad      $^2$Department of Mathematics, Shanghai Maritime University,   Shanghai,
  China.   $^*$\,\textsf{liulh@shmtu.edu.cn}    }

\date{}
\maketitle

\begin{quotation}
\textbf{Abstract.}  The primitive of private broadcast encryption introduced by Barth, Boneh and Waters,
 is used to encrypt a message to several recipients while hiding the identities of the recipients.
In their construction, a recipient has to first decrypt the received ciphertext to extract the verification key for one-time signature.
He then uses the verification key  to check whether the ciphertext is malformed.
The authors did not consider that information delivered over a channel, especially over a broadcast channel, should be authenticated as to its origin.
We remark that the conventional public key signature  suffices to authenticate data origin  and filter out all malformed ciphertexts.
 We also discuss the disadvantages of the primitive of one-time signature used in their construction.

\end{quotation}

\section{Introduction}
The primitive of broadcast encryption was formalized by Fiat and Naor \cite{FN93}, which requires that the broadcaster encrypts a message such that a particular
set of users can decrypt the message sent over a broadcast channel.
The Fiat-Naor broadcast encryption and the following works \cite{GSY99,GSW00,KRS99,S97,ST98} use a combinatorial
approach. This approach  has to right the  balance between the efficiency and the number of colluders that the system is resistant to.
Recently, Boneh et al \cite{BGW05,GW09} have constructed some broadcast encrypt systems. In these systems, the public parameters must be updated to allow more users.

In 2006,  Barth, Boneh and Waters \cite{BBW06} put forth a new cryptographic primitive, private broadcast encryption,
which is used to encrypt a message to several recipients while hiding the identities of the recipients. The primitive has many applications. For example,
commercial sites  can use it to protect identities of customers because
competitors might use this information for targeted advertising.
Their  construction \cite{BBW06}  is secure against an active attacker
 while achieving good efficiency.
 In the construction, a recipient has to first decrypt the received ciphertext to extract the verification key for one-time signature.
He then uses the verification key  to check whether the ciphertext is malformed.
The authors did not consider that information delivered over a channel, especially over a broadcast channel, should be authenticated as to its origin.

In this paper, we remark that the construction is impractical because the data origin authentication is very important to a broadcast encryption.
In real life, a listener must first authenticate the identity of the broadcaster.
It is unwise to decrypt the received message without authenticating its origin.
We also remark that the conventional public key signature suffices to authenticate data origin  and filter out all malformed ciphertexts.
Besides, we  discuss the disadvantages of the primitive of one-time signature used in their construction.

\section{Barth-Boneh-Waters Private Broadcast Encryption Scheme}
The private broadcast encryption scheme uses a public key encryption that has key indistinguishability
under CCA attacks (IK-CCA) to encrypt the ciphertext component for each recipient.
It then generates a random signature and verification key for a one-time, strongly
unforgeable signature scheme \cite{L79,R90}. It includes
the verification key in each public key encryption and then signs the entire ciphertext with
the signing key.

Suppose that (\textsf{Init}, \textsf{Gen}, \textsf{Enc}, \textsf{Dec}) is  a strongly-correct  IK-CCA public key scheme,  (\textsf{Sig-Gen}, \textsf{Sig}, \textsf{Ver}) is a strongly existentially
unforgeable signature scheme, and $(E, D)$ are  semantically secure symmetric key encryption
and decryption algorithms.  The private broadcast encryption system can be described as follows.

\textsf{Setup}($\lambda$): Given a security parameter $\lambda$, it generates global parameters $I$ for the system. Return \textsf{Init}($\lambda$).

\textsf{Keygen}($I$): Given the global parameters $I$, it generates public-secret key pairs.
For each user $i$, run $(pk_i, sk_i)\leftarrow \mbox{Gen}(I)$, return $(pk_i, sk_i)$ and publish $pk_i$.

\textsf{Encrypt}($S, M$): Given a set of public keys $S = {pk_1, \cdots, pk_n}$ generated by Keygen($I$)
and a message $M$, it generates a ciphertext $C$.

\hspace*{2mm} 1. $(vk, sk)\leftarrow  \mbox{\textsf{Sig-Gen}}(\lambda)$.

\hspace*{2mm} 2. Choose a random symmetric key $K$.

\hspace*{2mm} 3. For each $pk_i$ $\in S$, $c_{i}\leftarrow \mbox{\textsf{Enc}}_{pk_i}(vk||K)$.

\hspace*{2mm} 4. Let $C_1$ be the concatenation of the $c_{i}$, in random order.

\hspace*{2mm} 5. $C_2 \leftarrow E_K(M)$.

\hspace*{2mm} 6. $\sigma\leftarrow  \mbox{\textsf{Sig}}_{sk}(C_1||C_2)$.

\hspace*{2mm} 7. Return the ciphertext $C = \sigma||C_1||C_2$.

\textsf{Decrypt}($sk_i, C$): Given a ciphertext $C$ and a secret key $sk_i$, return $M$ if the corresponding
public key $pk_i \in S$, where $S$ is the set used to generate $C$. Decrypt can also return $\bot $ if $pk_i \notin S$
or if $C$ is malformed.
User $i$ parses $C$ as $\sigma||C_1||C_2$ and $C_1 = c_1|| \cdots||c_n$. For each $j \in \{1, \cdots, n\}$,

\hspace*{2mm} 1. $p \leftarrow \mbox{\textsf{Dec}}(sk_i, c_j)$.

\hspace*{2mm} 2. If $p$ is $\bot$, then continue to the next $j$.

\hspace*{2mm} 3. Otherwise, parse $p$ as $vk||K$.

\hspace*{2mm} 4. If $\mbox{\textsf{Ver}}_{vk}(C_1||C_2, \sigma)$, return $M = D_K(C_2)$.

\section{Drawbacks in the Barth-Boneh-Waters scheme}

The Barth-Boneh-Waters broadcast encryption scheme can provide recipient privacy. But we find the construction  has four drawbacks.
\begin{itemize}
\item[1.]
   The Barth-Boneh-Waters private  scheme assigns a pair of keys $(pk_i, sk_i)$ to each recipient $i$ for public key encryption and decryption.
   But it does not assign a pair of keys $(pk_B, sk_B)$ to the broadcaster for public key signature.
   The drawback leads the authors  not to  specify the procedure of data origin authentication.
   A valid recipient can  extract the verification key to authenticate data origin only after he successfully completes the procedure of public key decryption.
   That is to say, they adopt the strategy of data origin authentication after public key decryption.

We here stress that the construction is impractical because  the data origin authentication is very important to a broadcast encryption. In real life, a listener must first authenticate the identity of the broadcaster.
The listener then decides whether to decrypt the broadcasted message or not. It is unwise to decrypt the received message without authenticating its origin.
Indeed, the problem of SPAM is getting more and more serious, which has greatly affected our normal daily life and the public communication environment.

\item[2.]
The main purpose of introducing the one-time signature in their construction is to ensure that {an adversary cannot extract a ciphertext
component from the challenge ciphertext and use it in another ciphertext because it will be unable  to sign the new ciphertext under the same verification key}.
Simply speaking, its purpose is to filter out the malformed ciphertexts. We should  stress that
the conventional public key signature  suffices to authenticate data origin  and filter out the malformed ciphertexts.
It is unnecessary to introduce another mechanism to check the malformed ciphertexts.

\item[3.] The scheme specifies that the algorithm of Encrypt generates a random signature and verification key for a one-time, strongly
unforgeable signature scheme. It means that the broadcaster binds his identity with the verification key by himself, not by a trusted third party.
The description is incorrect.
 We stress that the verification key for signature, even for one-time signature, must be authenticated by a trusted third party.
It should be easily accessible and publicly available to the verifier.
Otherwise, the signature scheme is vulnerable to man-in-the-middle attack.

\item[4.] In the original scheme, the sign $\bot$ is unspecified. Thus, each recipient can not decide which $c_j$ is intended for him. Only after $vk$ is derived successfully and  the verification  $\mbox{\textsf{Ver}}_{vk}(C_1||C_2, \sigma)$ passes, he can decide. This incurs more cost because the recipient has to do the same number of public key decryptions $\mbox{\textsf{Dec}}(sk_i, c_j)$  as that of verifications $\mbox{\textsf{Ver}}_{vk}(C_1||C_2, \sigma)$.

\end{itemize}

\section{An improvement of the Barth-Boneh-Waters scheme}

In this section, we propose an improvement of the Barth-Boneh-Waters scheme. See Table 1 for its full description.

\vspace*{3mm}

\hspace*{25mm}   Table 1: The Barth-Boneh-Waters scheme and its improvement\vspace*{3mm}

\begin{tabular}{|l|l|}
  \hline
 The Barth-Boneh-Waters scheme & The improvement \\ \hline
 \textsf{Keygen}($I$): For each user $i$,  return & \textsf{Keygen}($I$): For each user $i$, return \\
   $(pk_i, sk_i)$ for public key encryption, &   $(pk_i, sk_i)$ for public key encryption, \\
   and publish $pk_i$. &  and publish $pk_i$.\\
  & For the broadcaster, \underline{return $(pk_B, sk_B)$}\\
  & for public key signature and publish $pk_B$.\\
  \hline
  \textsf{Encrypt}($S, M$): \uwave{$(vk, sk)\leftarrow  \mbox{\textsf{Sig-Gen}}(\lambda)$}. & \textsf{Encrypt}($S, M$): \\
     Choose a random symmetric key $K$.&  Choose a random symmetric key $K$.\\
 For each $pk_i$ $\in S$, \uwave{$c_{i}\leftarrow \mbox{\textsf{Enc}}_{pk_i}(vk||K)$}. &  For each $pk_i$ $\in S$, \underline{$c_{i}\leftarrow \mbox{\textsf{Enc}}_{pk_i}(pk_B||K)$}.\\
 Let $C_1$ be the concatenation of the $c_{i}$,& Let $C_1$ be the concatenation of the $c_{i}$,\\
 in random order.&  in random order.\\
 $C_2 \leftarrow E_K(M)$, \uwave{$\sigma\leftarrow  \mbox{\textsf{Sig}}_{sk}(C_1||C_2)$}. &  $C_2 \leftarrow E_K(M)$,  \underline{$\sigma\leftarrow  \mbox{\textsf{Sig}}_{sk_B}(C_1||C_2)$}. \\
   Return $C = \sigma||C_1||C_2$.&   Return $C = \sigma||C_1||C_2$.\\
 \hline
  \textsf{Decrypt}($sk_i, C$): User $i$ parses $C$ as $\sigma||C_1||C_2$& \textsf{Decrypt}($sk_i, C$): User $i$ parses $C$ as $\sigma||C_1||C_2$\\
   and $C_1 = c_1|| \cdots||c_n$.  &   and $C_1 = c_1|| \cdots||c_n$.  \\
 &  \underline{If $\mbox{\textsf{Ver}}_{pk_B}(C_1||C_2, \sigma)$ fails, return 0.} Otherwise,\\
 For each $j \in \{1, \cdots, n\}$, $p \leftarrow \mbox{\textsf{Dec}}(sk_i, c_j)$.  &  for each $j \in \{1, \cdots, n\}$, $p \leftarrow \mbox{\textsf{Dec}}(sk_i, c_j)$.\\
\uwave{If $p$ is $\bot$}, then continue to the next $j$. &   Parse $p$ as $pk'_B||K'$. If $pk'_B\neq pk_B$,\\
 Otherwise, parse $p$ as $vk||K$.& then continue to the next $j$. \\
\uwave{If $\mbox{\textsf{Ver}}_{vk}(C_1||C_2, \sigma)$}, return $M = D_K(C_2)$. &  Otherwise,  return $M' = D_{K'}(C_2)$.\\
  \hline
\end{tabular}\vspace*{3mm}

The basic idea behind the improvement is to assign a pair of keys $(pk_B, sk_B)$ to the broadcaster for public key signature.
The setting makes it possible for  a recipient to  authenticate data origin first of all. If it succeeds,
he then proceeds to the public key decryption. The strategy can greatly reduce a recipient's computational cost because it successfully filters out all origin-unknown
and malformed ciphertexts.

The Encryption algorithm computes
$$c_{i}= \mbox{\textsf{Enc}}_{pk_i}(pk_B||K)$$ for each $pk_i$ $\in S$. The added header $pk_B$ helps each user $i$ to
decide which component of the ciphertext is intended for himself because the probability of that the header of $\mbox{\textsf{Dec}}(sk_i, c_j)$ equals to the header of $\mbox{\textsf{Dec}}(sk_i, c_k)$ is negligible, where $j\neq k$.
Note that in the original scheme the sign $\bot$ is unspecified.  For convenience, we suggest to introduce the header $pk_B$ for checking.

The main difference between the original scheme and its improvement is that data origin authentication
must come before public key decryption. In the original Decryption algorithm, a user $i$ has to complete the procedure of public key decryption at first. He then extracts the verification key for one-time signature to filter out malformed ciphertexts. As we pointed out before, the setting results in that the original scheme is vulnerable to man-in-the-middle attack because the user $i$ does not access to the verification key through proper channels.
  To resist this trivial attack, we adopt the mechanism of public key signature instead of one-time signature.

\section{Remarks on one-time signature}

The primitive of one-time signature was invented by Leslie Lamport \cite{L79} in 1979.
Each Lamport public key can only be used to sign one single message, which means many keys have to be published if many messages are to be signed.
A hash tree can be used on those public keys, publishing the top hash of the hash tree instead. But this increases the size of the resulting signature because parts of the hash tree have to be included in the signature.
The researchers are familiar with one-time signature scheme presented by Merkle  \cite{M87},
 which is based on one-way functions, as opposed to trapdoor functions that are used in public key signatures.
Bleichenbacher and Maurer \cite{BM94, BM96} had
suggested one-time signatures based on acyclic graphs.

One-time signatures have been considered to be impractical because of complicated key management and long signature size.
Merkle \cite{M87, M89}  introduced the method of tree authentication to alleviate the problem of key management for a large number of one-time signatures.
Rohatgi \cite{R99} proposed some techniques to reduce the signature size.
Perrig \cite{P01} introduced  hash chains
for key management.
 Reyzin and Reyzin \cite{RR02} introduced a one-time signature scheme that has faster signature and verification times (for a
single signature). This scheme was improved by Pieprzyk et al. \cite{PWX03}.
The recent works of \cite{BKN04,JLM03,S04} have
improved Merkle's hash-tree method.
The one-time signature presented by Zaverucha and Stinson \cite{ZS11}
requires that PK size is of $O(\kappa n)$ bits, where $\kappa $ is the DL security parameter and $n$ is the number of bits in the
message to sign.
Naor et al. \cite{NSW05} suggest that when fast signatures are required, some one-time signatures
can be a promising alternative to the public-key signatures.

Although these one-time signatures are interesting, we would like to stress that the problem of efficient key management for one-time signatures still remains open.
This is due to that the cost to
guarantee the authenticity of a user's public key is expensive in
the scenario of Public Key Infrastructure (PKI for short). In nature, PKI entails that a user's public key should be
repeatedly usable in the life duration. This means the primitive of one-time signature is somewhat incompatible with PKI.

The conventional public key signatures are claimed to be vulnerable to quantum computers.
 But the performances of current quantum computers, D-Wave One and D-Wave Two, mitigate the threat.
 In May 2014, researchers \cite{SSSV14} at UC Berkeley and IBM  published a classical model
 explaining the D-Wave machine's observed behavior, suggesting that it may not be a quantum computer.
Any predictions on quantum computers have become more uncertain since the announcements of D-Wave systems.
 In the current situation, we think that it is unnecessary to use one-time signatures to replace conventional public key signatures.

\section{Conclusion}
In this paper we present an improvement of the Barth-Boneh-Waters  private broadcast encryption.
We also discuss the disadvantages of one-time signature used in their construction and stress that the primitive is inappropriate for a
broadcast system because of its complicated key management and long signature size.


\end{document}